\documentstyle[12pt,aaspp4,tighten]{article}

\input{psfig}

\def\VEV#1{\left\langle #1\right\rangle}

\def\la{\hbox{ \raise.35ex\rlap{$<$}\lower.6ex\hbox{$\sim$}\ }}
\def\ga{\hbox{ \raise.35ex\rlap{$>$}\lower.6ex\hbox{$\sim$}\ }}

\def\vecx{{\bf x}}
\def\veck{{\bf k}}
\def\W2{{\cal W}}

\slugcomment{\vbox{\hbox{CAL-663}
             \hbox{CU-TP-904}
             \hbox{astro-ph/9807211}}}

\begin{document}

\title{Weakly Nonlinear Clustering for Arbitrary Expansion Histories}

\author{Marc Kamionkowski$^*$\altaffilmark{1} and Ari
     Buchalter$^\dagger$\altaffilmark{2}}
\affil{$^*$Department of Physics, Columbia University, 538 West
     120th Street, New York NY 10027}
\affil{$^\dagger$Department of Astronomy, Columbia University,
     538 West 120th Street, New York NY 10027}
\altaffiltext{1}{kamion@phys.columbia.edu}
\altaffiltext{2}{ari@astro.columbia.edu}

\begin{abstract}
Bouchet et al. (1992) showed that in an open or closed Universe
with only pressureless matter, gravitational instability from
Gaussian initial conditions induces a normalized skewness, $S_3
\equiv \VEV{\delta^3} \VEV{\delta^2}^{-2}$, that has only a very
weak dependence on the nonrelativistic-matter density.  Here we
generalize this result to a plethora of models with various
contributions to the total energy density, including
nonrelativistic matter, a cosmological constant, and other forms
of missing energy.  Our numerical results show that the
skewness (and bispectrum) depend only very weakly ($\lesssim2\%$)
on the expansion history.  Thus, the skewness and bispectrum
provide a robust test of gravitational instability from Gaussian
initial conditions, independent of the underlying cosmological
model.
\end{abstract}


\section{INTRODUCTION}

One of the robust predictions of inflation that distinguishes it
{}from numerous alternatives such as topological-defect or
isocurvature models is that large-scale structure
grew from a Gaussian distribution of primordial density
perturbations.  However, even if the primordial distribution was
Gaussian, subsequent gravitational evolution would
introduce deviations from Gaussianity.  A Gaussian distribution
is symmetric with respect to overdense and underdense regions,
but as gravitational amplification of density perturbations
proceeds, overdensities can become arbitrarily large.  On the
other hand, any given region can become only so underdense
(since densities must be positive).  Thus, the resulting
distribution is necessarily skewed.

Peebles (1980) calculated perturbatively the normalized skewness 
induced by gravitational instability of initially Gaussian
perturbations  in an Einstein-de Sitter universe.  More
precisely, if $\delta(\vecx)\equiv 
[\rho(\vecx)-\bar\rho]/\bar\rho$ is the fractional density 
perturbation and $\rho(\vecx)$ is the density 
at position $\vecx$, then in an Einstein-de Sitter universe, the 
statistic $S_3 \equiv \VEV{\delta^3} \VEV{\delta^2}^{-2}=34/7$.
Bouchet et al. (1992) generalized this result to an open or
closed universe containing only pressureless matter and found
that the prediction for $S_3$ remains the same to $< 2\%$
for any reasonable value of $\Omega_0>0.1$.  Thus, they showed
that measurements of $S_3$ from galaxy surveys should provide a
robust test of Gaussian initial conditions, independent of
$\Omega_0$, but only if the universe consists of just
pressureless matter.

However, slow-roll inflation predicts that the universe is flat,
while observations seem to point to a matter density $\Omega_0
\simeq 0.2-0.5 < 1$.  Therefore, if inflation is correct, then
there must be some other missing energy density.  The most
widely considered form for this density is a cosmological
constant (with pressure given by $p = -\rho$), but theorists
have also realized that this missing energy might be
characterized by any of a number of equations of state.
Among these are some energy density that
scales as $a^{-2}$ ($K$-matter), where $a(t)$ is the scale factor 
of the universe (\cite{kol89}; \cite{kam96}; \cite{pen97}), something akin
to an evolving cosmological constant, possibly generated by the
dynamics of some scalar field (\cite{cob97}; \cite{sil97};
\cite{tur97}; \cite{cal98}), or some combination of
several forms of missing energy.  

To test for primordial Gaussianity of density perturbations in
this cornucopia of missing-energy models, we calculate the
induced skewness in flat models with a cosmological constant
and/or some other form(s) of energy.  For completeness we also
consider open and closed universes with a variety of equations
of state.  For all cosmological models investigated,
the numerical results for $S_3$ never deviate from 34/7 by more
than 2\% for models with $\Omega_0 \geq 0.1$, and no more than
about $1\%$ for $\Omega_0\geq0.3$.  Thus, we conclude that the
normalized skewness does indeed provide a robust test for
primordial Gaussianity, independent of the nature of the species
that constitute the total energy density.

The bispectrum (Fourier transform of the induced 
three-point correlation function) 
has also been calculated, as a byproduct of the calculation
of the skewness, for Gaussian initial
conditions (\cite{pee80}; \cite{fry84}; \cite{gor86}) in an
Einstein-de Sitter universe and in a universe with pressureless
matter (\cite{bou92}).  Here we provide results for the
bispectrum for the models we investigate.

The details of our calculation are presented in the next Section.  
We write
down the differential equation, for an arbitrary expansion history, 
that must be solved to obtain the growth of the skewness.  
In Section 3, we provide numerical 
results for the skewness and bispectrum for several forms of missing energy and
for some models with various combinations of missing energy.  We 
also provide an analytic approximation to these quantities as a function of 
$\Omega_0$ for flat cosmological-constant models.  We then make
some concluding remarks.

\section{CALCULATION}

The calculation of the bispectrum and skewness is outlined in
detail in Peebles (1980) and Fry (1984), and we simply highlight
the relevant steps below; Bouchet et al. (1992) employ an
alternative approach based on the Zeldovich approximation.
The Friedmann equation for the expansion rate 
as a function of redshift, $z$, is given by
\begin{equation}
     H(z) \equiv {\dot a\over a} = H_0 E(z); \;\;\;\;\;\;\;\; 
     E(z) = \sqrt{\sum_i \Omega_i (1+z)^{3(1+w_i)}},
\label{eq:expansionrate}
\end{equation}
where $H_0$ is the present-day value of the
Hubble parameter, the dot denotes a derivative with respect
to time $t$, and $\sum_i \Omega_i = 1$ where $\Omega_i$ represents
the contribution to the overall energy density from species $i$ having
equation of state $p=w_i \rho$.  In this formulation, the curvature
of the universe contributes an amount $\Omega_K$ to
the total energy density, and yields the term $\Omega_K (1+z)^2$
in the sum in equation (\ref{eq:expansionrate}).
Thus, in a universe with, for example, nonrelativistic-matter 
density $\Omega_0$ and cosmological-constant contribution to
closure density of $\Omega_\Lambda$, $E(z)=\sqrt{\Omega_0(1+z)^3 
+ \Omega_\Lambda + (1-\Omega-\Omega_\Lambda)(1+z)^2}$.  The
deceleration can be written as
\begin{equation}
     {\ddot a \over a} =H_0^2 F(z),
\label{eq:deceleration}
\end{equation}
and $F(z)$ can be obtained by differentiating
equation (\ref{eq:expansionrate}).  For example, for the $E(z)$ given 
above, $F(z)=\Omega_\Lambda-\Omega_0(1+z)^3/2$.

If the fractional density perturbation is small ($\delta
\ll 1$), the equations of motion for $\delta(\vecx,t)$ can be
solved perturbatively using the expansion $\delta(\vecx,t) =
\delta^{(1)}(\vecx,t) + \delta^{(2)}(\vecx,t) + \cdots$, with
$\delta^{(i+1)} \ll \delta^{(i)}$.  The perturbative equations
for the lowest-order term, $\delta^{(1)}$, turn out to be
separable, $\delta^{(1)}(\vecx,t)=\delta_1(\vecx)D_1(t)$, and the
familiar linear-theory growth factor satisfies
\begin{equation}
     \ddot D_1 +2{\dot a \over a} \dot D_1 - {3\over 2} \Omega_0
     H_0^2\left( {a_0 \over a} \right)^3 D_1=0.
\label{eq:lineartheory}
\end{equation}

The next-order equation of motion is
\begin{eqnarray}
     \ddot \delta^{(2)} &+& 2 {\dot a \over a} \dot \delta^{(2)} -
     {3 \over 2}
     \Omega_0 H_0^2 \left( {a_0 \over a} \right)^3 \delta^{(2)}
     \nonumber \\
     &=&
     \left[ {3 \over2}\Omega_0 
     H_0^2 \left({a_0 \over a} \right)^3 +\left( {\dot D_1\over
     D_1}\right)^2 \right] D_1^2 \delta_1^2
     \nonumber \\
     & + &    \left[{3 \Omega_0 H_0^2 \over 2} \left( {a_0 \over
     a} \right)^3 +2 \left( {\dot D_1 \over D_1} \right)^2
     \right] D_1^2 \delta_{1,i} \Delta_{1,i} + \left( {\dot D_1
     \over D_1} \right)^2 D_1^2
     (\Delta_{1,ij})  (\Delta_{1,ij}),
\label{eq:longequation}
\end{eqnarray}
where
\begin{equation}
     \Delta_1(\vecx) = - {1\over 4\pi} \int \, d^3x' \,
     {\delta_1(\vecx) \over |\vecx-\vecx'|}.
\label{eq:Deltaequation}
\end{equation}
The homogeneous part of
equation (\ref{eq:longequation}) is the same as that for
equation (\ref{eq:lineartheory}), and it turns out that the solution
for $\delta^{(2)}$ is dominated by the inhomogeneous terms on the
right-hand side.  Since the inhomogeneous part is the sum of
three terms with two different time dependences, the solution for
$\delta^{(2)}$ is not obviously separable.  However, the complete
solution is simply the sum of the solutions to the equations given by
the homogeneous equation driven by each of these sources
separately.  We may thus write the solution as
\begin{equation}
     \delta^{(2)} = (D_{2,a}+D_{2,b}) \delta_1^2 +
     (D_{2,a}+2D_{2,b}) \delta_{1,i} (\Delta_{1,i}) + D_{2,b}
     \Delta_{1,ij} \Delta_{1,ij},
\label{eq:solution}
\end{equation}
where the second-order growth factors satisfy
\begin{equation}
     \ddot D_{2,a} +2{\dot a \over a} \dot D_{2,a} - 
     {3\over 2}\Omega_0 H_0^2 \left({a_0 \over a} \right)^3
     D_{2,a} = {3\over 2} \Omega_0 H_0^2 \left({a_0 \over a}
     \right)^3 D_1^2,
\label{eq:D2aequation}
\end{equation}
\begin{equation}
     \ddot D_{2,b} +2{\dot a \over a} \dot D_{2,b} - {3 \over
     2} \Omega_0 H_0^2 \left({a_0 \over a} \right)^3
     D_{2,b} = \dot D_1^2,
\label{eq:D2bequation}
\end{equation}
with the initial conditions that the growth factors and their
first time derivatives are zero at $t=0$.  Although it appears
that there are two differential equations (for $D_{2,a}$ and
$D_{2,b}$) to be solved for the second-order solution, it can
be verified that $D_{2,b}=(D_1^2-D_{2,a})/2$.  Thus, there
is really only one independent second-order
differential equation.  The first of these
equations (\ref{eq:D2aequation}) can be
solved analytically for the Einstein-de Sitter model with the
result that $D_1 \propto t^{2/3}$ and $D_{2,a}=(3/7)D_1^2$.
For more general models, equations (\ref{eq:lineartheory}) and
(\ref{eq:D2aequation}) can be integrated numerically.  

It is then straightforward to follow the arguments given in
\S 18 of Peebles (1980) to find the normalized skewness.
In terms of the quantity $\mu=D_{2,a}/D_1^2$ [this is twice
the parameter $\kappa$ in Bouchet et al. (1992)], it is given by
\begin{equation}
     S_3 = 4 + 2\mu = \frac{34}{7} + \frac{6}{7}\left(\frac{7}{3}\mu -1 \right)
\label{eq:S3equation}
\end{equation}
We have thus enumerated a simple prescription for calculating
$S_3$ in a model with Gaussian initial conditions and an
arbitrary expansion history.

In terms of the parameter $\mu$, the second-order correction to 
the density contrast is (\cite{bou92}),
\begin{equation}
     \delta^{(2)} = {D_1^2\over 2} \left[(1+\mu) \delta_1^2 + 2
     \delta_{1,i} \Delta_{1,i} + (1-\mu) \Delta_{1,ij}
     \Delta_{1,ij} \right].
\label{eq:delta2equation}
\end{equation}
If we compare equation (11) in Fry (1984) with equation (A2) in
Goroff et al. (1986), it is clear that the unsymmetrized scaled
bispectrum for arbitrary expansion history can be written 
[cf. equation (A1) in Goroff et al. (1986)]
\begin{equation}
     P_2^{(s)}(\veck_1,\veck_2)={1\over 2} \left[(1+\mu) +
     {\veck_1 \cdot \veck_2 \over k_1 k_2} \left( {k_1 \over k_2}
     + {k_2 \over 
     k_1} \right) + (1-\mu) \left( { \veck_1 \cdot \veck_2 \over 
     k_1 k_2} \right)^2 \right].
\label{eq:bispectrum}
\end{equation}

\section{RESULTS}

\begin{figure}
\plotone{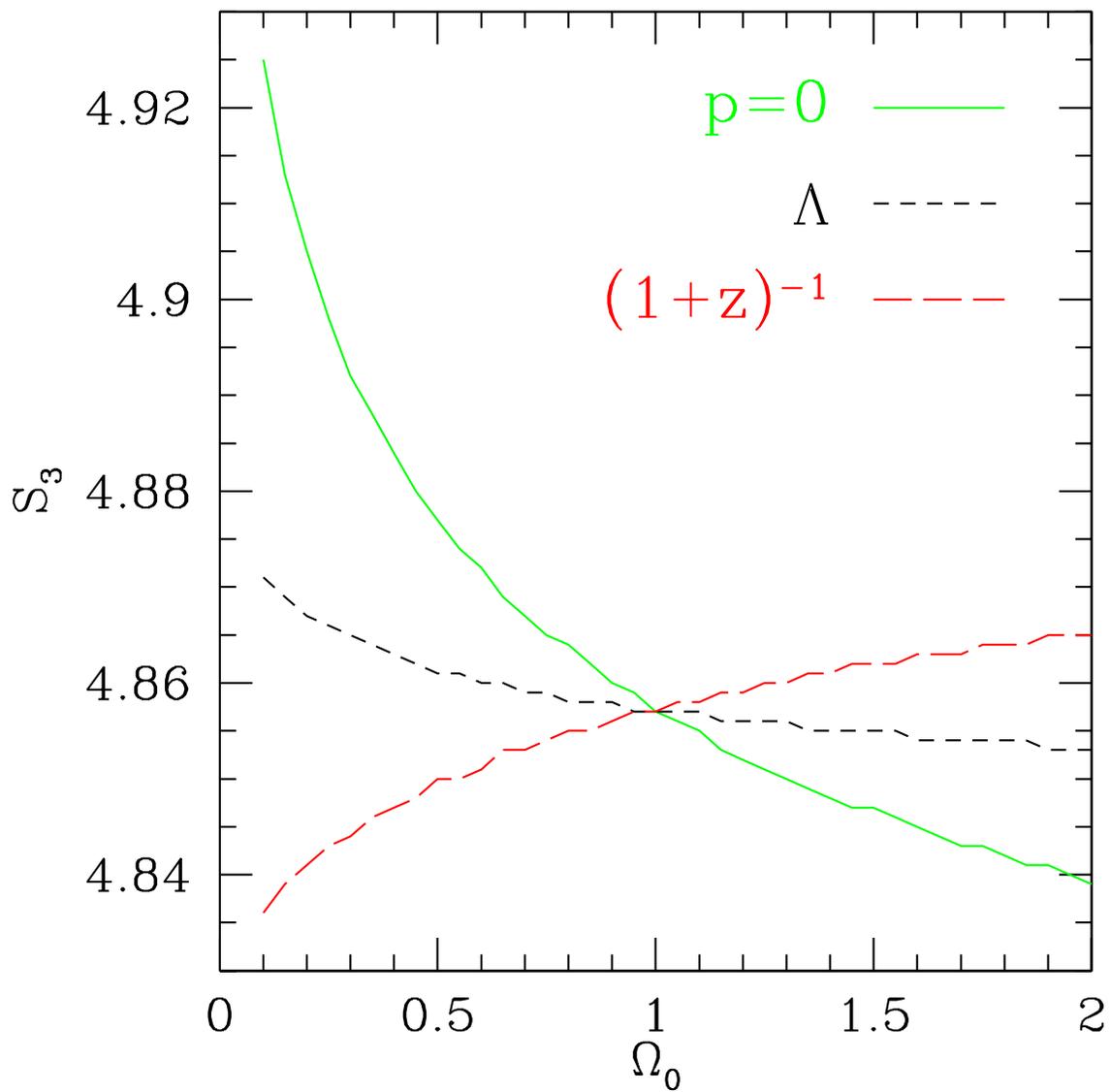}
\caption{The normalized skewness $S_3$ as a function of the
     nonrelativistic-matter density $\Omega_0$ in a universe
     with only nonrelativistic matter (solid curve), a flat
     cosmological-constant model (short-dash curve), and a flat
     model with an energy density that scales as $(1+z)^{-1}$
     (long-dash curve).}
\label{fig:S3plot}
\end{figure}

The solid curve in Fig. \ref{fig:S3plot} shows the normalized
skewness $S_3$ as a function of the nonrelativistic-matter
density $\Omega_0$ in an open, flat, or closed model containing
only pressureless matter; i.e., in a 
model with $E^2(z)=\Omega_0(1+z)^3+(1-\Omega_0)(1+z)^2$.  [To
clarify, the results do not really depend on the geometry 
of the universe, but rather on the expansion history
$E(z)$.]  This reproduces the result of Bouchet et al. (1992)
who found that the quantity $\mu$ is well approximated for
$0.05<\Omega<3$ by
\begin{equation}
     \mu \simeq {3 \over 7}\Omega_0^{-2/63}.
\label{eq:bouchet}
\end{equation} 

The short-dashed curve in Fig. \ref{fig:S3plot} shows our
results for the normalized skewness $S_3$ as a function of the 
nonrelativistic-matter density $\Omega_0$ in a flat 
cosmological-constant model\footnote{Bouchet et al. (1995) have 
also integrated these equations numerically for the
cosmological-constant universe}; that is,
$E^2(z)=\Omega_0(1+z)^3+(1-\Omega_0)$ (note that $\Lambda < 0$
for models with $\Omega_0>0$).  We see
that the difference between $S_3$ and 34/7 is even smaller in a
cosmological-constant model than in an open model with the same
$\Omega_0$.  We find that $\mu$ is well approximated in these
models by
\begin{equation}
     \mu \simeq {3 \over 7} \Omega_0^{-1/140}.
\label{eq:lambdaresult}
\end{equation}

To generalize even further, the long-dashed curve in
Fig. \ref{fig:S3plot} shows the normalized skewness $S_3$ in a flat
model with a nonrelativistic-matter density $\Omega_0$ plus some
hypothetical matter density with $p=-2\rho/3$, scaling as 
$(1+z)^{-1}$; that is,
$E^2(z)=\Omega_0(1+z)^3+(1-\Omega_0)(1+z)$ (we will refer to
this as ``$\Omega_1$-matter'' below).  Again, the
correction to $S_3$ is very small for any reasonable value of
$\Omega_0$.  The normalized skewness is well approximated in these
models by
\begin{equation} 
     S_3 \simeq {3\over 7} \Omega_0^{1/77}.
\label{eq:omegaoneresult}
\end{equation}

A flat model with nonrelativistic matter and some $K$-matter
density, such as nonintersecting cosmic strings, which scales as
$(1+z)^{-2}$ (\cite{kol89}; \cite{kam96}; \cite{pen97}), has
the same $E(z)$ as the standard open model with the same
nonrelativistic-matter density (since $w_K=-1/3$).  Thus, the
skewness will be given by the solid curve in Fig. \ref{fig:S3plot}.

Of course, the cases
shown in the Figure do not exhaust the full range of $E(z)$, and
we can think of no way to systematically study all possible
cases.  However, we have tried numerous strange combinations of
the types of matter considered above, and in no case do we find
any deviation from 34/7 significantly larger than that in the
open model.  Several examples are shown in Table 1.  In
addition to flat $\Omega_0<1$ models with various
combinations of cosmological constant, $K$-matter, and
$\Omega_1$-matter (including some with a negative cosmological constant),
we also consider a closed cosmological-constant model
that is just consistent with quasar-lensing statistics
(\cite{whi96}) and an $\Omega_0=10$ closed model (\cite{har93}).  We
include these models to demonstrate the robustness of $S_3$ to
dramatic variations in the expansion history.  We do not suggest 
that all of these models are observationally tenable.
These numerical experiments lead us to believe with
good confidence that $S_3$ must differ from 34/7 by no more than
2\% in any cosmological model with $\Omega_0 > 0.1$, and to no
more than $\simeq1\%$ for any model with $\Omega_0\geq0.3$, if
structure grew via gravitational instability from a Gaussian
distribution of primordial density perturbations.

\begin{table*}[tb]
\begin{center}
\begin{tabular}{|c|c|c|c|c|c|c|c|}
\hline
 $\Omega_0$ & $\Omega_\Lambda$ & $\Omega_K$ & $\Omega_1$ & 
       $S_3$ & \% $S_3$ & $\mu$ \\ \hline \hline
 1  &  0  &  0  &  0  &  4.857  &  0 & 0.429  \\ \hline
 0.3  &  0.7  &  0  &  0  &  4.865  &  0.15 & 0.432  \\ \hline
 0.3  &  0  &  0.7  &  0  &  4.892  &  0.73 & 0.446 \\ \hline
 0.3  &  0  &  0  &  0.7  &  4.844  &  -0.26 & 0.422 \\ \hline
 0.3  &  0.35  &  0.35  &  0  &  4.880  &  0.48 & 0.440 \\ \hline
 0.3  &  0.35  &  0  &  0.35  &  4.854  &  -0.07 & 0.427 \\ \hline
 0.3  &  0 &  0.35  &  0.35  &  4.872  &  0.31 & 0.436 \\ \hline
 0.3  &  -1  &  1.7  &  0  &  4.915  &  1.19  & 0.458 \\ \hline
 0.3  &  -1  &  0  &  1.7  &  4.824  & -0.67 & 0.412 \\ \hline
 10  & 0  &  -9  &  0  &  4.809  & -0.99  & 0.405 \\ \hline
 2  & 2  &  -3  &  0  &  4.796  & -1.25  & 0.398 \\ \hline
\end{tabular}
\end{center}
\caption{Predicted skewnesses and $\mu$.  The second to last
     column lists the percentage deviation of $S_3$ from 34/7.}
\label{tab:predictions}
\end{table*}

\section{DISCUSSION}

We have generalized the result of Bouchet et al. (1992)---that
the normalized skewness and bispectrum are 
only very weakly sensitive to $\Omega_0$ in models with
only pressureless matter---to general Friedmann cosmologies with
arbitrary contributions to the energy density, including 
the simplest cosmological-constant model.  The robustness
of these measures allows for excellent model-independent tests
of structure formation from gravitational instability of a
Gaussian distribution of primordial perturbations.

We have only considered the effect of the expansion on the
skewness and bispectrum.  One might wonder whether perturbations in the
scalar field in models with a dynamical scalar field might
affect the growth of structure.  If so, then it is conceivable
that the skewness would deviate from the
gravitational-instability prediction.  However, structure
formation in this case would not be due simply to gravitational
instability, as we have assumed here.  Analytic approximations
suggest that on small scales ($\lesssim100\,h^{-1}$ Mpc),
scalar-field ordering should not significantly affect the
skewness (\cite{jaf94}).  Thus we expect that in (non-topological) 
scalar-field models, the prediction $S_3 \simeq34/7$ should hold 
for Gaussian initial conditions.

In practice, other factors must be considered when comparing
measurements of the observed skewness and bispectrum with the
predictions above.  For example, the skewness calculated
above is that for an unsmoothed density field, while the
observed distribution is intrinsically discrete.  Bernardeau
(1994) considered the effects of smoothing, and it is
straightforward to apply the discussion therein to generalize
our results for $S_3$.  In particular, for the case of smoothing
with a spherical top-hat filter for scale-free power spectra
$P(k) \propto k^n$,
\begin{equation}
     S_{3,s} = S_3 - (3+n),
\end{equation}
where $S_{3,s}$ is the result for the smoothed distribution.
In addition, the skewness and bispectrum presented here are those 
for the underlying mass distribution.  Galaxy surveys probe the
luminous-matter distribution, which if biased relative to the
matter, will be characterized by a different skewness and
bispectrum.  If the galaxy fractional density
perturbation is written as
\begin{equation}
    \delta_g(\vecx,t) = b_1 \delta(\vecx,t) + \frac{b_2}{2}
    \delta^2(\vecx,t) + \cdots, 
\label{eq:dexp}
\end{equation}
where $b_1$ is the linear bias term, $b_2$ the first nonlinear
term, and so on, then the skewness of the observed galaxy
distribution is given by (\cite{frg93}) 
\begin{equation}
     S_{3,g} = \frac{1}{b_1}S_{3,s} + \frac{3 b_2}{b_1^2},
\end{equation}
and similar corrections have been obtained for the bispectrum
(\cite{fry94}).  More realistic calculations involving, for
example, true scale-dependent power spectra or bias evolution
have been obtained using numerical techniques (\cite{fry96};
\cite{jib97}; \cite{gab98}; \cite{buc98}).
The model-independence of the predicted skewness and bispectrum
for the matter distribution makes determinations of the bias
{}from measurements of these quantities (\cite{mat97};
\cite{buc98}) that much more robust.

The robustness of the scaled skewness to the expansion history
supports the notion that the geometry of gravitationally
collapsing objects is determined almost exclusively by the
initial conditions.  This leads us to speculate that the
predictions for the higher-order moments and correlation
functions (e.g., kurtosis, four-point correlation
function, trispectrum, and so forth) will be similarly
independent of the expansion history.  It would be interesting
and straightforward to check this hypothesis numerically.

\bigskip
This work was supported by D.O.E. contract DEFG02-92-ER 40699, NASA
NAG5-3091, NSF AST94-19906, and the Alfred P. Sloan Foundation.

{}


\begin{thebibliography}{}

\bibitem[Bernardeau 1994]{ber94} Bernardeau, F. 1994, A\&A, 291, 697

\bibitem[Bouchet et al. 1992]{bou92} Bouchet, F. R., et al.
     1992, ApJ, 394, L5

\bibitem[Bouchet et al. 1995]{bou95} Bouchet, F. R., et al. 1995,
     A\&A, 296, 575

\bibitem[Buchalter \& Kamionkowski 1998]{buc98} Buchalter,
     A. \& Kamionkowski, M. 1998, in preparation

\bibitem[Caldwell, Dave, \& Steinhardt 1998]{cal98} Caldwell,
     R. R., Dave, R., \& Steinhardt, P. J. 1998,
     Phys. Rev. Lett., 80, 1582

\bibitem[Coble, Dodelson, \& Frieman 1997]{cob97} Coble, K.,
     Dodelson, S., \& Frieman, J. 1997, Phys. Rev. D, 55, 1851

\bibitem[Fry 1984]{fry84} Fry, J. N. 1984, ApJ, 279, 499

\bibitem[Fry 1994]{fry94} Fry, J. N. 1994, Phys. Rev. Lett., 73,
     215

\bibitem[Fry 1996]{fry96} Fry, J. N. 1996, ApJ, 461, L65

\bibitem[Fry \& Gazta\~{n}aga 1993]{frg93} Fry, J. N. \& Gazta\~{n}aga, E. 
     1993, ApJ, 413, 447

\bibitem[Gazta\~{n}aga \& Bernardeau 1998]{gab98}  Gazta\~{n}aga, E. \&
     Bernardeau, F. 1998, A\&A, 331, 829

\bibitem[Goroff et al. 1986]{gor86} Goroff, M. H., et al. 1986,
     ApJ, 311, 6

\bibitem[Harrison 1993]{har93} Harrison, E. 1993, ApJ, 405, L1

\bibitem[Jaffe 1994]{jaf94} Jaffe, A. H. 1994, Phys. Rev. D, 49, 
     3893

\bibitem[Jing \& Borner]{jib97} Jing, Y. \& B\"{o}rner, G. 1997, 
     A\&A, 318, 667

\bibitem[Kamionkowski \& Toumbas 1996]{kam96} Kamionkowski,
     M. \& Toumbas, N. 1996, Phys. Rev. Lett., 77, 587

\bibitem[Kolb 1989]{kol89} Kolb, E. W. 1989, ApJ, 344, 543

\bibitem[Matarrese, Verde, \& Heavens 1997]{mat97} Matarrese,
     S., Verde, L., Heavens, A. F. 1997, MNRAS, 290, 651

\bibitem[Peebles 1980]{pee80} Peebles, P. J. E. 1980, The
     Large-Scale Structure of the Universe (Princeton
     Univ. Press)

\bibitem[Pen \& Spergel 1997]{pen97} Pen, U.-L. \& Spergel,
     D. N. 1997, ApJ, 491, L67

\bibitem[Silveira \& Waga 1997]{sil97} Silveira, V. \& Waga,
     I. 1997, Phys. Rev. D, 56, 4626

\bibitem[Turner \& White 1997]{tur97} Turner, M. S. \& White,
     M. 1997, Phys. Rev. D, 56, 4439

\bibitem[White \& Scott 1996]{whi96} White, M. \& Scott,
     D. 1996, ApJ, 459, 415

\end{thebibliography}
\end{document}